\documentstyle{article}

\newcommand{\mathbf}{\bf}

\begin{document}

\begin{center}
{\huge\bf On the Stability of Yang-Mills Bundles over $S^4$}
\end{center}

\vspace{1cm}
\begin{center}
{\large\bf 
F.GHABOUSSI}\\
\end{center}

\begin{center}
\begin{minipage}{8cm}
Department of Physics, University of Konstanz\\
P.O. Box 5560, D 78434 Konstanz, Germany\\
E-mail: ghabousi@kaluza.physik.uni-konstanz.de
\end{minipage}
\end{center}

\vspace{1cm}

\begin{center}
{\large{\bf Abstract}}
\end{center}
The stability of Yang-Mills bundles over the usual $S^4$ space-time manifold is investigated according to the topological methods. The necessary gauge- and topological invaraint criterion for the exsitence of the related critical points is defined. It is shown that according to this criterion there exists no critical point even for the action functional of the standard U(1) gauge theory of electrodynamics on a $S^4$ manifold in view of its topological structure and therefore such a theory can not be stable. We will discuss also a general consequence of this result according to which for a stable U(1) Yang-Mills theory over a compact 4-manifold, this manifold should possess some self consistent compact 2-manifold substructure. These results are also in agreement with the known very general result for the {\it structural stability} of dynamical systems.
\begin{center}
\begin{minipage}{12cm}

\end{minipage}
\end{center}

\newpage
{\Large 1. Introduction and summary}

The mathematical theories of three of the four fundamental interactions of physics are commonly considered as Yang-Mills field theories on fibre bundles with unitary gauge groups over $S^4$ which represents the compact space-time base manifold. Here we investigate the stability of such theories in the standard mathematical manner \cite{morssm}. We consider the stability of the critical points of the action functional of Maxwell theory of electrodynamics which is given as a Yang-Mills functional on a U(1) fibre bundle over $S^4$ manifold. The main question is whether such critical points which possess certain topological properties, exist on a fibre bundles with $S^4$ base manifold, or that already their existence requires structural conditions on the topology of the base manifold of bundles. Our following results show that the desired stability is not given for the $S^4$ case, but it can be given if the base manifold possesses compact 2-dimensional structure $(\sim S^2)$. Although these results can be surprising, nevertheless they agree not only with the well known general results for the {\it stuructural stability} of dynamical systems \cite{morssm} and the previous positive results of Atiyah and Bott for similar cases ($\sim S^2$), but they agree also with previous negative results of C. H. Taubes for the $S^4$ case (see below).

In view of the fact that the stability of the critical points of functionals requires the {\it existence} of non-degenerate critical points, where the critical points of a Yang-Mills functional on a bundle are given as Yang-Mills connections \cite{bourg}. Further since connections are subject to gauge transformations by which such connections are produced locally, therefore one has to look for some gauge invariant criterion or measure for their existence. Indeed there is such a measure for the U(1) connections which is given by the gauge- and even topological invariant {\it harmonic} curvature two form of these connections \cite{abactz}. Then the {\it main gauge invariant property} of U(1) Yang-Mills connections is that they possess harmonic Yang-Mills curvature 2-forms. Therefore the {\it existence} and the stability of such critical points on a $(M, U(1))$ bundle over a simply connected compact orientable riemannian 4-manifold (M), requires the existence of harmonic Yang-Mills 2-forms $Harm^2 (M, U(1))$ on this bundle manifold \cite{bourg}, \cite{nak}. Note that these harmonic curvature forms are the gauge invariant measure of the U(1) Yang-Mills connections, therefore the existence of such harmonic curvature forms on a bundle ensure also the existence of the critical points of Yang-Mills connections type on the same bundle. Further recall that since in  the case of $U(1)$ fibre, the $Harm^2 (M, U(1))$ on the bundle manifold is isomorphic to the $Harm^2 (M)$ on the base manifold, then the existence of such critical points of Yang-Mills connection type requires also the existence of $Harm^2 (M)$ on the base manifold. 
Furthermore since according to the Hodge theorem $Harm^2 (M)$ are isomorphic to the $H^2 (M)$ elements of the second cohomology group on $M$ \cite{nak},  \cite{uone}; therefore the existence and the stability of such critical points requires equally the existence of $H^2 (M)$ according to the both isomorphisms: $Harm^2 (M, U(1)) \cong Harm^2 (M) \cong H^2 (M) \cong H^2 (M, U(1))$ on the above mentioned simply connected 4-manifold. Note that in view of this requirement of the existence of $H^2 (M)$ for the exitence of such critical points of Yang-Mills connection type, there can be no such critical points for the case where: ($M = S^4$), because $H^2 (S^4) = 0$ by definition \cite{nak}. However concerning the isomorphism $Harm^2 (M, U(1)) \cong H^2 (M, U(1))$ the main question is again whether there exists $H^2 (S^4, U(1))$ on such a bundle with the $S^4$ base manifold in order to ensure the existence of the desired critical points of Yang-Mills connection type. 

Nevertheless note that in general, on the one hand a Yang-Mills functional on a bundle manifold with a certain fibre manifold $F$ over the base manifold $B$, is defined as a Yang-Mills functional on the whole bundle manifold: (B, F). On the other hand the existence of cohomology groups or their dual homology groups are ensured on bundle manifolds, in general, by the spectral sequence method \cite{span}. Thus there is also the Kuenneth formula for the existence of co/homology groups in the case where the bundle is a product bundle $(B, F) := B \times F$. Therefore to ensure the existence of the critical points of Yang-Mills connection type for functionals on the (B, U(1)) bundle, one has to look in general at the {\it existence} of the required $H^2 (B, U(1))$ according to these methods. The necessity of application of topological methodes in this work is motivated by the facts that firstly the gauge independent/invaraint measure for the existence of critical points of Yang-Mills connection type is given by the differential topologic quantity: $Harm^2 (B, U(1))$. Secondly the global existence of this quantity is given by the global topologic quantity: $b^2 := dim (Harm^2 (B, U(1)))$, which ensures the global existence of such critical points and the necessary global stability of the theory on the space-time manifold: $B$. 

We will investigate here in the first part both possibilities, i. e. the Kuenneth formula and the spectral sequence method, for the most important case of the Maxwell theory on a compact orientable riemannian space-time 4-manifold, where $B := S^4$ and $F := U(1) = S^1$. It is shown that since there is no second cohomology group class on this bundle manifold, i. e. since $H^2 (S^4, S^1) = 0$, therefore there is also no isomorphic second harmonic form on this bundle: $Harm^2 (S^4, S^1) = 0$. Then there can be no such critical points of Yang-Mills connection type for the $U(1)$-Yang-Mills functional (YM) over $S^4$, in view of the above mentioned result according to which the critical points of this functional should possess $Harm^2 (S^4, S^1)$ curvature forms \cite{abactz}, \cite{bourg}. In other words the desired critical points are absent for the case where the base manifold is $S^4$, since the gauge invariant measure for their existence, i. e. $Harm^2 (S^4, S^1) \cong H^2 (S^4, S^1)$ is absent. If so, then we have a major problem with the stability of the most basic gauge theory of physics over $S^4$. We show further  that this problem can be solved only if one assumes a compact space-time 4-manifold $M$ {\it with a compact two dimensional substructure} ($\sim S^2 \in M$) in view of the fact that ($S^2 \in M$) admits the necessary $H^2 (M)$ structure for the above mentioned existence criterion. The proper reason for this circumstance is that {\it by definition} the $S^4$ manifold does not possess any second homology class $H_2 (S^4)$ and hence also its dual class $H^2 (S^4)$ according to the de Rham theorem; whereas the $S^2$ manifold possesses both classes by definition \cite{nak}. Furthermore we show that the existence of the mentioned harmonic curvature form for the Yang-Mills connections ensures also the stability requirement according to which the second variation of the functional should be positive.

It is very importent to mention that in view of the duality of $H^2 (M)$ and $H_2 (M)$ according to the de Rham theorem, this existence result about the necessity of some $H^2 (M)$ or its dual $( S^2 \in H_2 (M))$ homology substructure for space-time follows also from the geoemetry of the integral Maxwell equations which are integrals over some closed 2-manifolds ($\sim S^2 \in H_2 (M)$) (see below). In other words our result according to the stability and existence of critical points of the fibre bundle theory of electrodynamics is in best agreement with the geoemtrical structure of the integral Maxwell equations and one should achive the same result, if one considers this geometrical structure carefully. 

Note that there were a lot of topological studies of Yang-Mills bundles over compact manifolds and their stability aspects in the last few decades, notably those in the works of Atiyah and Bott. Nevertheless the acheived {\it certain results} on the stability of critical points of Yang-Mills-functionals, are those on bundles over {\it 2-dimensional} "Riemann surfaces" only, e. g.  $S^2$ \cite{atias}, which is in best agreement with our positive results for manifolds with compact 2-dimensional struture. Whereas as these authors noticed the stability of critical points of Yang-Mills-functionals over $S^4$ could not be achived and remained for the future \cite{atias}. These hopes however were not fulfilled, but rejected by the later results. These further results showed that there are no stable Yang-Mills bundles over the $S^4$ manifold \cite{tacsi}, but there {\it can be} some {\it possibly} stable Yang-Mills bundles which are constructed as {\it new} bundles from a Yang-Mills bundle over $S^4$. These are either {\it new} bundles with base manifolds which are constructed by removing some "points" from $S^4$ base manifold \cite{sed}, or {\it new} bundles which are constructed from the original Yang-Mills bundle over $S^4$ by gluing technics \cite{taczdv}. In other words our negative result for the case $S^4$ is also in best agreement with these results. 

The paper is organized as follows: we will show in the Sect. 2 that by the help of differential geometric methods the variation can be achived directly in a very compact and fundamental form. In Sect. 3 we obtain the above mentioned negative results with respect to the stability of critical points of a $U(1)$ functional on bundles over $S^4$, with the help of algebraic topological methods very directly and then in Sect. 4 we will make an estimate that only the Yang-Mills bundles over compact 4-manifolds with some $S^2$ substructure, e. g. $S^2 \times S^2$, may possess the desired stable critical points. The Sect. 5 is devoted to conclusions and perspectives.

{\Large 2. The differential geometry and topology of variation}

First note that one can verify the harmonicity of the curvature forms of critical points of Yang-Mills connection type for a U(1)-Yang-Mills functional on a compact orientable riemannian 4-manifold (YM (4D)) with the help of differential geometry directly without explicite variation. Herefore one should consider this funcional as a zero form $\omega^0 (4D)$ and the related Yang-Mills-Lagrangian density as a $U(1)$ 4-form $(F \wedge * F) \in \omega^4 (4D)$ where $F \in \omega^2 (4D)$ and $* F$ is its Hodge dual 2-form:

\begin{equation}
\omega^0 (4D, U(1)) \sim \omega^0 (4D) = \int_{4D} \omega^4 (4D) = \int_{4D} F \wedge * F,  
\end{equation}

where we omit the remarks on U(1) and (4D) from now on some times.
The critical points of this functional are given according to: $\delta \omega^0 = d \omega^0 =0$, since in view of the identity $d^{\dagger} \omega^0 \equiv 0$, the only non-trivial variation of a zero-form should be given by: $d \omega^0  = 0$. Further note that the $ (\omega^4)$ is the highest form on the bundle manifold and should automatically obey: $d \omega^4  \equiv 0$. 

Then we have by the Hodge duality $\omega^0 = * \omega^4 = * (F \wedge * F)$ on a compact orientable riemannian 4-manifold (4D), the following relations from the variation principle of the $\omega^0 (4D)$ functional: 

\begin{equation}
d \omega^0 = d * (F \wedge * F) = 0, d^{\dagger} (F \wedge * F) = 0,  
\end{equation}

and: 

\begin{equation}
d^{\dagger} \omega^0 \equiv d^{\dagger} * (F \wedge * F) = 0,  d (F \wedge * F) \equiv 0,  
\end{equation}

from the above identity, where $d^{\dagger} := * d *$.

Hence we obtain for the {\it critical points} of the U(1) Yang-Mills functional directly:

\begin{equation}
d^{\dagger} F (4D) = 0, d F (4D) \equiv 0, 
\end{equation}

where $d F (4D) \equiv 0$ is an identity which is equivalent to the above mentioned identity $d^{\dagger} \omega^0 (4D) \equiv 0$. This is the same result which is achived if one considers the varaition of the above action functional with respect to the U(1) connection 1-form $A$, where:
$F := d A$.
 
It is known from the Hodge theory that the 2-form $F$ satisfying equations (4) is a harmonic 2-form: $F \in Harm^2 (4D)$. Then the curvature form of the critical points of the U(1) Yang-Mills-functional, i. e. the curvature form of the Yang-Mills connections, are given as the harmonic 2-form: $F \in Harm^2 (4D)$. Therefore the true gauge invariant exitence of these critical points requires the existence of related harmonic curvature 2-forms: $Harm^2 (4D) \cong Harm^2 (4D, U(1))$ on the bundle manifold, which are isomorphic to: $H^2 (4D, U(1))$. In other words, in order that the existence of these critical points is ensured in a gauge invariant manner, there should be a $H^2 (4D, U(1))$ class on the bundle manifold. In the case under consideration where the base manifold of bundle is $S^4$ and the fibre is $U(1) = S^1$, we have to prove that whether the 2-form $H^2 (S^4, S^1)$ exists or not. 

Note here that, there is $H^2$ neither on the base manifold $S^4$, nor on the fibre manifold $S^1$, in view of the fact that there are only $H^l$ and $H^0$ on a sphere manifold $S^l$ and all other co/homology classes vanish: $H^a (S^l) = 0, a = 1,...,l -1$. Hence {\it in the first sight} it is natural that there is no $H^2$ also on the bundle manifold the bundle manifold $(S^4, S^1)$.
 Nevertheless there are cases where a co/homology group like $H^2$ is present on a bundle manifold like $(S^1 \times S^1)$, although it is present neither on the base manifold $S^1$ nor on the fibre manifold $S^1$. Hence the absence of $H^2$ on $(S^4, S^1)$, is a highly non-trivial case, even so that $H^2 (S^4) \sim H^2 (S^1) = 0$. Secondly note that, as we will show, both the spectral sequence method and the Kuenneth formula result in the absence of $H^2 (S^4, S^1)$ and $H^2 (S^4 \times S^1)$ respectively,  in agreement with each other. So that there is no possibility to avoid this absence. 

{\Large 3. The existence conditions for co/homologies on bundles}

To show the absence of $H^2$ in the mentioned cases, first we apply the Kuenneth formula for the direct product bundle: $M = (S^4 \times S^1)$.

This formula is given for $M = (B \times F)$ by:

\begin{equation}
H^a (B \times F) = \oplus_{p+q = a} [ H^p (B) \otimes H^q (F) ],  
\end{equation}

where $a$ determines the existing co/homology classes on the product bundle under consideration. For the case of sphere bundles with $B = S^n$ and $F = S^m$ we obtain:

\begin{equation}
H^a (S^n \times S^m) = \oplus_{p+q = a} \ [ H^p (S^n) \otimes H^q (S^m)] 
\end{equation}

Nevertheless in view of the restriction of non-vanishing co/homology groups on sphere manifolds $S^l$ to $H^0 (S^l)$, $H^l (S^l)$, it is obvious that the only non-zero co/homology groups in this case are given for:

\begin{equation}
H^a (M = (S^n \times S^m)) \neq 0, iff: a = 0, m, n, ( m + n )
\end{equation}

Hence in our case we obtain:

\begin{equation}
H^a (S^4 \times S^1) = \oplus_{p+q = a} [ H^p (S^4) \otimes H^q (S^1)], 
\end{equation}

which is given in term of the Betti numbers by: 

\begin{equation}
b^a (S^4 \times S^1) = \Sigma_{p+q = a} \ b^p (S^4) \cdot b^q (S^1) 
\end{equation}

Then for the desired case of $a = 2$ we obtain:

\begin{equation}
H^2 (S^4 \times S^1) = \oplus_{p+q = a} [ H^p (S^4) \otimes H^q (S^1) ] = H^0 (S^4) \otimes H^2 (S^1) \oplus H^1 (S^4) \otimes H^1 (S^1) \oplus H^2 (S^4) \otimes H^0 (S^1) = 0, 
\end{equation}

where also the positivity of the rank of co/homology groups is used.

Accordingly the related Betti numbers formula also gives a vanishing result for $H^2 (S^4 \times S^1)$:

\begin{equation}
b^2 (S^4 \times S^1) = \Sigma_{p+q = 2} \ b^p (S^4) \cdot b^q (S^1) = b^0 (S^4) \cdot  b^2 (S^1) + b^1 (S^4)  \cdot  b^1 (S^1) + b^2 (S^4) \cdot  b^0 (S^1) = 0
\end{equation}

Obviously this is so, since $H^2 (S^4) \sim H^2 (S^1) \sim H^1 (S^4) \sim H^1 (S^1) = 0$. 

Hence there is no $H^2 (S^4 \times S^1)$ and also no $Harm^2 (S^4 \times S^1)$ according to the Hodge theorem. Therefore there can be no Yang-Mills connections with harmonic curvature forms, as the desired critical points of a Yang-Mills functional on a {\it product bundle} with $U(1)$ fibres over $S^4$.

To consider the spectral sequence method in our case, note that the spectral sequence method should be considered as a {\it generalization} of Kuenneth formula for general bundles. Thus, in principle, this method should give results for a general bundle $(B, F)$, in accordance with that of Kuenneth formula for the product bundle where $(B, F) = (B \times F)$. In other words, it should be expected that if the Kuenneth formula results in a vanishing co/homology class on a product bundle with certain base- and fibre manifolds, then also the spectral sequence method will result in the vanishing of the same co/homology class on a general bundle with the same base- and fibre manifold.

The spectral sequence of relevant cohomology classes is given by the following {\it simplified version} of exact sequence formula on the fibre bundle manifold $(B, F)$ where $B$ the basis manifold is a cohomology $n$-sphere over {\mathbf R} \cite{span}:

\begin{equation}
... \rightarrow H^{a-1} (B, F) \rightarrow H^{a-1} (F) \rightarrow H^{a-n} (F) \rightarrow H^a (B, F) \rightarrow H^a (F) \rightarrow ...,
\end{equation}

where the periodicity of the sequence is obvious. Here the index $a$ determines again the relevant co/homology classes on the $(B, F)$ bundle with fibre manifold $F$, where the related homology sequence would be directed in the {\it opposite} direction $... \leftarrow ...$ according to the {\it dual} action of the boundary operator.

Then for the case where $B = S^n$ and $F = S^m$ we obtain 

\begin{equation}
... \rightarrow H^{a-1} (S^n, S^m) \rightarrow H^{a-1} (S^m) \rightarrow H^{a-n} (S^m) \rightarrow H^a (S^n, S^m) \rightarrow H^a (S^m) \rightarrow ...,
\end{equation}

The existence of a certain co/homology group on the bundle manifold $(B, F)$, e. g. $H^a (B, F)$, implies that at least one of its neighbour co/homology groups should be non-zero. In other words, in order that $H^a (B, F)$ exists, it is necessary that at least either $H^{a-n} (F)$, or that $H^a (F)$ is non-zero.

The calculation for the "sphere bundles", i. e. where both the base manifold and the fibre manifold are spheres, is simplified, in view of the mentioned restriction of non-vanishing co/homology groups on sphere manifolds $S^l$ to the $H^0 (S^l)$ and $H^l (S^l)$ co/homology groups. In other words we have the following algebraic relation for the existence of a certain co/homology group $H^a (S^n, S^m)$ on these Sphere bundles:

\begin{equation}
H^a (S^n, S^m) \neq 0, iff: (a - n = 0, m), (a = 0, m), or  \  iff: (a = 0, m, n, (m + n)),
\end{equation}

where also the positivity of the rank of co/homology groups is used. Note the obvious agreement of this relation for the spectral sequence method with the Kuenneth formula for the product sphere bundles where the mentioned restriction of co/homology groups is the same as in the relation (7).

Hence we obtain for the $ (S^4, S^1) $ bundle of electrodynamics:

\begin{equation}
H^a (S^4, S^1) \neq 0, iff: a = 0, 1, 4, 5,
\end{equation}

as the ranks of the existing co/homology group classes.

Accordingly, since the rank of the desired cohomology group for the stability, $H^2 (S^4, S^1)$, is absent in relation (15) which requires according to the Hodge theorem that also $Harm^2 (S^4, S^1) = 0$. Then also according to the spectral sequence method, there are no Yang-Mills connections with harmonic curvature form as the critical points of a $U(1)$ Yang-Mills functional on a bundle over $S^4$.

We conclude that in view of the triviality of the second Betti number for a $U(1)$ bundle with a $S^4$ base manifold, i. e. $b^2 (S^4, S^1) = 0$ which is equal to the dimension of the related second harmonic forms: $dim  (Harm^2 (S^4, S^1))$, the Yang-Mills-functional which is defined on this bundle does not possess any critical points of Yang-Mills connection type, i. e. with harmonic curvature forms.

These results reject the usefulness of $S^4$ as a $4D$ base manifold of a stable gauge theory of electrodynamics. In other words only those 4-manifolds seems to be good in this respect, which possess self consistent, 2-dimensional, orientable, compact submanifolds, in view of the fact that only $U(1)$ bundles with these base manifolds admit the required $H^2$ class. 

Now we show that the desired existence and hence also the desired existence and stability of critical points of the U(1) Yang-Mills-functional on bundles over 4-manifolds (4D) is given only for those cases where the base manifold possesses such two dimensional submanifolds, e. g. $(4 D) = (S^2 \times S^2)$:

{\Large 4. Manifolds which admit stable electrodynamics}

We saw from the above discussions that the reason why there are no critical points of Yang-Mills connection type for the Yang-Mills-functional on a $U(1)$-bundle over $S^4$ base manifold, is that neither $S^4$, nor the $U(1) \cong S^1$ fibre manifold possess $H^2$ cohomology class. In other words, in order that the bundle manifold possesses $H^2 \cong Harm^2$ which is required by the exitence of these critical points, at least either the base- or the fibre manifold should possess the $H^2$ cohomology class. Then the other possibility where both base and fibre manifolds are of $S^1$ type, is rejected in the case of the 4-dimensional base manifold by definition.
Hence in view of the fact that $S^1$ fibre manifold can not possess the desired $H^2$ by definition, it is necessary that the base manifold should possess a $H^2$ cohomology class. But this is only given, if the base manifold possesses some closed 2-manifold which possesses such a $H^2$ class. We mentioned also that the proper reason for this circumstance is that {\it by definition} the $S^4$ manifold does not possess any second homology class $H_2 (S^4)$ and hence also its dual class $H^2 (S^4)$ according to the de Rham theorem. Then among the euclidean spheres $S^l$ it is only the $S^2$ manifold which possesses a $H_2$ class and hence also its dual class $H^2$ by definition \cite{nak}. In other words, since the invariant measure of critical points of Yang-Mills connection type is given by $F \in Harm^2$ classes on the bundle and base manifolds, which are isomorphic to $H^2$ classes there on. Therefore according to the isomorphism between such $H^2$ classes and $H_2$ classes on the same manifolds, the invaraint existence of such critical points requires the existence of $H_2$ class on the base or bundle manifold. Then the topological invaraint measure of any curvature like ($F \in Harm^2 \in H^2$), i. e. the second Betti number $b^2 = b_2$, can be given by the integral of ($F \in H^2$) over a certain compact manifold: $\sim S^2 \in H_2$ \cite{nak}. This fact is related with that which is known also from the integral Maxwell equations which are integrals of electromagnetic curvature $F$ over some closed two dimensional manifolds $\sim S^2$ or over some closed curves on these manifolds (see below). Then to admit a stable theory of electrodynamics in a topological invariant form, the base manifold should admit some compact two dimensional substructure which supports such topological requirements. As an example of such a suitable structure one may consider $S^2 \times S^2$ as the base 4-manifold of the U(1) fibre bundle, since in this case the base manifold possesses even two $H^2 (S^2 \times S^2)$ classes according to the above discussed Kuenneth formula: 

\begin{equation}
H^2 (S^2 \times S^2) = H^0 (S^2) \otimes H^2 (S^2) \oplus H^2 (S^2) \otimes H^0 (S^2) = 2 H^0 (S^2) \cdot H^2 (S^2) 
\end{equation}

Then such a fibre bundle $(( S^2 \times S^2) \times S^1)$ also possesses a $Harm^2 ((S^2 \times S^2)  \times S^1 ) \cong H^2 ((S^2 \times S^2)  \times S^1)$ class according to the same formula and the desired existence and the stability of Yang-Mills connections on this fibre bundle is ensured. Note also that in view of the above mentioned dual exact homology sequences \cite{span} and the duality between $H^2 (M)$ and $H_2 (M)$ \cite{nak}, one obtains the homological version of the above result according to which a stable $U(1)$ bundle over $S^4$ can not be stable, because the absence of $H_2 (S^4, S^1)$ results in the absence of $H^2 (S^4, S^1)$. 
Consequently in order that a $U(1)$ bundle over a 4-manifold $M$ possess the desired critical points, this manifold should possesses some $H_2 (M)$ substructure, where $S^2 \in H_2 (M)$. In this manner also the exact homology sequence methode results in the same necessity of some $S^2$ substructure for the physical space-time.

{\Large 5. Conclusions and perspectives}

In conclusion note first that one may consider in general the triviality of $b_2 (S^4, S^1) \sim b^2 (S^4, S^1) \sim b^2 (S^4)$ as the reason for the instability of Yang-Mills fields on this bundle. Further note that if the 4-manifold is of $(\sim S^2 \times S^2)$ type, then in view of the existence of ($Harm^2 (S^2, S^1) > 0 \sim Harm^2 (S^2) > 0$) as the positive curvature form of the desired U(1) Yang-Mills connections on the base manifold $S^2$, one can obtain from the second variation of the functional: 

\begin{equation}
\omega^0 (S^2, S^1) := \int \int_{S^2} Harm^2 (S^2, S^1) \sim \int \int_{S^2} Harm^2 (S^2), 
\end{equation}

the value $Harm^2 (S^2) > 0$ which is a positive constant according to the constancy of the curvature of $S^2$. Therefore if one uses the suitable Yang-Mills functional on the 4-manifold $(\sim S^2 \times S^2)$ defined by: 

\begin{equation}
\omega^0 (S^2 \times S^2, S^1) := \int_{(S^2 \times S^2)} Harm^2 (S^2 \times S^2, S^1) \wedge 
Harm^2 (S^2 \times S^2, S^1), 
\end{equation}

which is {\it proportional} to:

\begin{equation}
\omega^0 (S^2 \times S^2) = \int_{(S^2 \times S^2)} Harm^2 (S^2 \times S^2) \wedge Harm^2 (S^2 \times S^2), 
\end{equation}

according to the above mentioned isomorphism between the topological quantities of the $S^1$- or $U(1)$ bundle manifold and its base manifold: $Harm^2 (M, S^1) \cong Harm^2 (M)$ 
\cite{uone}. Note that this integral is {\it proportional} to the product of integrals: 

\begin{equation}
\omega^0 (S^2) = \int_{S^2} Harm^2 (S^2) \cdot \int_{S^2} Harm^2 (S^2),
\end{equation}

according to the relations between the topological quantities of $(S^2 \times S^2)$ and $(S^2)$ manifolds , e. g. given by the relation (16), and further in view of the above mentioned isomorphism between hamonic forms and cohomology elements on a manifold: $Harm^2 (S^2) \cong H^2 (S^2)$. 

Then one obtains according to the above mentioned proportionalities also for the second variation of the above functional $\omega^0 (S^2 \times S^2, S^1)$ a positive constant value according to the above arguments and the stability requirement of the positivity of the second variation is fulfilled. In this manner both stability requirements, i. e. for the first and the second variations, could be fulfilled for the U(1) case, if the base space-time manifold possesses a $(\sim S^2 \times S^2)$ structure. Then the existence of the U(1) harmonic curvature form of the Yang-Mills connections which is proportional to the {\it positive and constant} curvature of $S^2$, ensures not only the {\it vanishing} of the first variation of the related U(1) functional, but it ensures also the {\it positivity} of the second variation of the same functional.

Note that on the one hand the above mentioned positive result of Atiyah and Bott with respect to the stability of Yang-Mills fields over compact  2-manifolds, e. g. $S^2$ \cite{atias}, underlines the correctness of our results. Then we showed above by very genral arguments that why the stability standards necessiate the presence of such a comapact 2-manifold. On the other hand the absence of the success in the case of $S^4$ there \cite{atias} and in further works, e. g. \cite{tacsi}, \cite{sed} and \cite{taczdv}, again underline also our results. Then as we showed above the compact 2-manifold structure is absent in $S^4$.

Secondly that the above achived negative result with respect to the critical points of Yang-Mills functional on the $U(1)$ fibre bundle over a $S^4$ base manifold can be achived, in principle, also according to the fact that in view of the above mentioned isomorphism \cite{uone}, according to which one has for this case: $H^2 (S^4) \cong H^2 (S^4, U(1)) \cong Harm^2 (S^4, U(1)) \cong Harm^2 (S^4)$ \cite{uone}. Then in view of the fact that $H^2 (S^4) \cong Harm^2 (S^4) \equiv 0$, one obtains also $H^2 (S^4, U(1)) \cong Harm^2 (S^4, U(1)) = 0$, which is the above obtained result by very general methods. However in order to achive the general results, one has to use the more general methods like those which are applied above. 

Thirdly one should look for a generalization of these results for the non-abelian Yang-Mills-bundles, since according to the Utiyama's lemma which is discussed by R. S. Palais \cite{palais}, the non-abelian Yang-Mills field strength $F = D A = d A + [A, A]$ is reduciable {\it  in the quasi canonical gauge} on some points on the base manifold of the bundle to its pure derivative term:  $F = d A$. {\it If so}, then according to this Palais-Utiyama construction, one could be free from difficulties which are related with a desired generalization of the Hodge theory for the non-abelian cases. Consequently the "covaraint harmonic" 2- form solution of the non-abelian equations of motion $(D F = 0, D^{\dagger} F = 0)$, is reduced according to this construction \cite{palais} to several components of the usual harmonic 2-form {\it on the base manifold}: i. e. to $(d F (M) = 0, d^{\dagger} F(M) = 0)$. Hence our result for the $U(1) \sim S^1$ case, should be generalizable at least for all those Yang-Mills-bundles with $SU(n), SO(n)$ fibres which are isomorphic to some spheres $S^n$, e. g. for $SU(2) \sim S^3$. In other words according to the mentioned Palais-Utiyama construction, the desired stability should be absent for all Yang-Mills-bundles with $S^4$ base manifold and $(\sim S^n, n \neq 2)$ fibres, in view of the absence of $H^2 \cong Harm^2$ on these bundles. Furthermore note that as we mentioned above, on the one hand the theory of stability requires the existence of a certain gauge invaraint quantity which is defined on the bundle manifold. But in the case of non-abelian Yang-Mills theories such quantities can be delivered only by a well defined differential topologic structure on the bundle manifold $(B, F)$, e. g.  according to a "generalized Hodge-de Rham theory" for non-abelian cases. On the other hand there exists no generalization of the Hodge-de Rham theory for cases where the $(B, F)$ manifold possesses non-abelian symmetries, i. e. there exists no such well defined differential geometrical set up including topological invaraints for non-abelian models of physical theories. Accordingly the stability of such non-abelian 4D theories is highly questionable from two sides: In general in view of the absence of any mathematical theory for non-abelian differential topology which can deliver an invaraint measure for the existence of critical points of any functional over any 4-dimensional base manifolds with {\it inherent} non-abelian symmetries. Specially in view of the absence of a similar mathematical theory for invaraint measures for the existence of critical points of Yang-Mills functionals on non-abelian fibre bundles over such 4-dimensional base manifolds. Then the usual Hodge-de Rham theory does not apply even on a 4-manifold with inherent non-abelian symmetry. Nevertheless as it is mentioned above, a possible solution to this problem could be, that in order to retain the standard stability theory for the physical Yang-Mills theories, one accepts some 2-dimensional submanifold structure with an inherent abelian symmetry for the physical space-time 4-manifold. 

There is another way to arrive at our above results about the preference of some 2-dimensional substructure for stable electrodynamics over 4-manifolds, namely according to the investigation of  the geometrical structure of the integral Maxwell equations. These are the integrals or solutions of the above mentioned differential equations of motion or those with sources $(d F = 0, d^{\dagger} F = J)$, which are given by integrals of electromagnetic 2-forms performed over some closed 2-manifolds ($\sim S^2$) or over some closed curves ($\sim S^1$). One may consider these closed curves ($\sim S^1$) as geodesic curves over the mentioned closed 2-manifolds ($\sim S^2$). In this manner the integral Maxwell equations also prefer some compact 2-manifold structure ($\sim S^2$) within the original space-time 4-manifold. The relation between this prefered ($\sim S^2$) submanifold by integral Maxwell equations and our results, is given by the homology version of our cohomological results. Then as we mentioned above, the homological structure of the stable $U(1)$ fibre bundle theory also requires also according to the spectral sequence method that the underlying 4-manifold $M$ should admit some $H_2 (M)$ homology, where ($\sim S^2 \in H_2 (M)$). These ($\sim S^2 \in H_2 (M)$) submanifolds can be considered as the above mentioned integration domain of the integral Maxwell equations {\it where these integral equations are defined at all}! Thus according to the above mentioned de Rham duality the topological invariant integrals of electromagnetic 2-forms which are elements of $H^2 (M)$, are performed over cycles which are elements of the $H_2 (M)$ \cite{nak}, where $H^2$ elements include also $harm^2$ elements. In this manner the well known physical facts about the Maxwell theory, i. e. the necessary {\it invariance} of the integral Maxwell equations, support our results with respect to the necessity of some 2-manifold substructure ($\sim S^2$) for space-time 4-mnifold. Thus the topological invaraint measure for the existence of $H^2 (M) \cong H_2 (M)$ is given by the non-triviality of $b^2 (M) = b_2 (M)$. In other words our result for the {\it topological invaraint measure} for the existence of critical points of U(1) Yang-Mills connection type, matches exactly with the physically required invaraince of  the integral Maxwell equations. A fact which underlines the physical relevance and correctnesw of our result.

Furthermore note that our result about the necessity of some 2-manifold of $S^2$ type for the stability of critical points of Yang-Mills connection type is also in agreement with results from the structural stability of dynamical systems according to which: only vector fields on compact 2-manifolds can be structurally stable whereas on higher dimensional compact manifolds one has no certainty for the stablity of vector fields, since there are always non-stable vector fields which disturbe their overall stability \cite{morssm}. Thus if the vector fields represent the above mentioned differential equations $d \omega^0 (4D) = 0, d ^{\dagger} \omega^0 = 0$ of the electrodynamical action functional, which are the Hodge duals of the differential euations: $d F (4D) = 0, d^{\dagger} F (4D) = 0$. Then according to the structural stability theorems \cite{morssm},  in order that these vector fields are {\it structurally stable}, the base manifold where these vector fields are defined should be of compact 2-manifold type. Then the stuructural stability of the above mentioned 4-dimensional vector fields is guaranted, {\it iff} the original 4-manifold (4D) possesses some compact 2-manifold substructure $((2D) \times (2D) ) \sim (S^2 \times S^2)$, where the above mentioned 4-dimensional vector fields are properly defined and they are reduced to: $d \omega^0 (2D) = 0, d ^{\dagger} \omega^0 (2D)  = 0$ with the Hodge duals: $d F (2D) = 0, d^{\dagger} F (2D) = 0$. In other words iff the 4-dimensional manifold where the vector fields are considered, possesses such a $((2D) \times (2D) ) \sim (S^2 \times S^2)$ structure, then these vector fields can be structurally stable iff they can be defined directly on the comapct $(2D)$ manifolds. This is but the same which results from our topological investigations that even if one defines the quantities of the Maxwell theory on a compact 4-manifold, nevertheless the stability of the critical point solutions of the theory requires that the 4-manifold should possesses some compact 2-manifold submanifolds $((2D) \sim S^2)$ where the critical points are actually defined. Thus the structural stability of a system is in principle the same as the global or topological stability of the system. In other words both structural stability or (our topological investigations) require the same property for the base manifold of stable dynamical systems: in order to have structurally  stable vector fields or (globally stable critical points of Yang-Mills connection type), these vector fields or (critical points) should be defined properly on compact 2-manifolds. Then it is the compact 2-manifold structure $((2D) \sim S^2)$ which ensures the structural stability of vector fields: $(\sim d \omega^0 (2D) = 0)$ or the topological stability of critical points: $(\sim F \in Harm^2 (2D))$, even if one thinks the compact 2D to be embedded in a 4D manifold: ($(2D) \in ((2D) \times (2D))$).

Note that our reslt concerning the preference of 4-dimensional base manifold with a substructure $(\sim S^2 \times S^2)$, is also in agreement with the previous results concerning the structure of simply connected 4-manifolds which possess similar substructures $(\sim S^2 \times S^2)$ \cite{fredm}. Also results concerning closed topological 4-manifolds show that the aboved favoured $(\sim S^2 \times S^2)$ substructure is the essential substructure of topological 4-manifolds \cite{nat}. This conicidence between our and the early topological results \cite{fredm}, \cite{nat}, underlines the correctness of our results which states the necessity of a compact 2D substructure for the physical 4-manifold. Thus it is known that the equations of electrodynamics are the best varification of the celebrated Hodge theory of harmonic integrals and topology.

It is important to mention a fundamental remark on smooth 4-manifolds, according to which: "At this moment there is no {\it existence} and uniqueness theorem for smooth 4- manifolds." \cite{kosinski}. Thus our results which reject $S^4$ as a useful manifold for the physical space-time, seems to be also in good agreement with the fact that smooth 4-manifolds {\it without substructures} like $(\sim S^2 \times S^2)$ are questionable at all in view of the absence of any existence theorem for them.

\bigskip 
{\Large Footnotes and References}


\begin{thebibliography}{100}
\bibitem{abactz}
These are solutions to the Maxwell equations $d F = 0, d^{\dagger} F = 0$ (see also Refs. \cite{nak} and \cite{atias}). We use here the terms of "stability of critical points of functionals" or the "stability of functionals" as synonyms.
\bibitem{uone}
Note that the Hodge-de Rham theory applies in the case of U(1) gauge theory of the electrodynamics in view of the commutativity of 
the U(1) algebra.
\end{thebibliography}
\end{document}